\documentclass[11pt]{article}

\usepackage{acl}
 \usepackage{tikz}
 \usetikzlibrary{calc,arrows.meta}
 \definecolor{corallight}{RGB}{250,236,231}   
 \definecolor{coraldark} {RGB}{153, 60, 29}   
 \definecolor{teallight} {RGB}{225,245,238}   
 \definecolor{tealmid}   {RGB}{ 93,202,165}   
 \definecolor{tealdark}  {RGB}{ 15,110, 86}   
\newcommand{\tyes}{\textcolor{tealdark}{\checkmark}}
\newcommand{\tno}{\textcolor{coraldark}{\ensuremath{\times}}}
\usepackage{times}
\usepackage{latexsym}
\usepackage[T1]{fontenc}
\usepackage[utf8]{inputenc}
\usepackage{enumitem}
\usepackage{microtype}
\usepackage{graphicx}
\usepackage[table]{xcolor}
\usetikzlibrary{arrows.meta, positioning, fit, calc}
\usepackage{placeins}
\usepackage{stfloats}
\usepackage{subcaption}
\usepackage{booktabs}
\usepackage{multirow}
\usepackage{tcolorbox}
\tcbset{
  fontupper=\footnotesize,
  boxsep=2pt,
  left=4pt,
  right=4pt,
  top=2pt,
  bottom=2pt,
  arc=1pt,
  boxrule=0.5pt,
  fonttitle=\footnotesize\bfseries
}
\usepackage{algorithm}
\usepackage{algorithmic}
\usepackage[nointegrals]{wasysym}

\usepackage{amsmath}
\usepackage{amssymb}
\usepackage{mathtools}
\usepackage{amsthm}

\usepackage[capitalize,noabbrev]{cleveref}

\theoremstyle{plain}

\theoremstyle{definition}

\theoremstyle{remark}

\usepackage[textsize=tiny]{todonotes}

\hypersetup{
  pdfkeywords={Jailbreak, LLM Safety, Black-box Attack, ECHO, Adversarial Attack}
}

\title{ECHO: Replacing the Compliance Proxy with Response Harmfulness in Optimization-Based Jailbreak Attack}

\author{
\textbf{Yuanwei Wu\textsuperscript{1}} \quad
\textbf{Lingdong Huang\textsuperscript{2}} \quad
\textbf{Yunlong Mao\textsuperscript{1,*}} \\
\textbf{Kanye Ye Wang\textsuperscript{3,*}} \quad
\textbf{Jingyu Hua\textsuperscript{1}} \quad
\textbf{Sheng Zhong\textsuperscript{1}} \\
\normalfont
\textsuperscript{1}Nanjing University \\
\textsuperscript{2}University of the Chinese Academy of Sciences \\
\textsuperscript{3}University of Macau \\
{\small \textsuperscript{*}\textbf{Correspondence:}
\texttt{maoyl@nju.edu.cn} ,
\texttt{wangye@um.edu.mo}}
}

\begin{document}
\maketitle

\begin{abstract}
Despite extensive safety alignment, large language models remain vulnerable to jailbreak attacks. Existing optimization-based attacks share a single objective: maximize a token-level proxy of victim compliance---such as the likelihood of an affirmative prefix---computed on the prompt rather than the decoded reply. This proxy cannot distinguish a cautious hedge from a fully actionable harmful answer, leaving the search blind to what the model actually says. We propose ECHO (Evaluation-Conditioned Harmfulness Optimization), which replaces the compliance proxy with response-level, continuous harmfulness scores from HarmSig, a 1.7B regressor grounded in actual harmful capability. ECHO couples HarmSig with five word-level perturbation operators and Keyword-Anchored Beam Search with global top-$m$ memory, under a strict black-box threat model. Across five open-weight victims on AdvBench, ECHO achieves up to 0.898 StrongREJECT mean score and 89.23\% HarmBench success. Our code is publicly available at \url{https://github.com/STANWYW/echo}.
\end{abstract}

\section{Introduction}
\label{sec:introduction}

Large language models (LLMs) are increasingly deployed in high-stakes products, yet they remain vulnerable to jailbreak attacks: adversarial prompts that bypass safety alignment and elicit disallowed content~\cite{zou2023universal,wei2023jailbroken}. Efficient jailbreak methods are crucial because they expose vulnerabilities in deployed APIs and supply adversarial prompts that downstream defenses such as adversarial fine-tuning, refusal training, and circuit breakers all consume at scale~\cite{krauss2025twinbreak,zhao2026unravelingllmjailbreakssafety}.

\begin{figure}[t]
\centering
\includegraphics[width=0.482\textwidth]{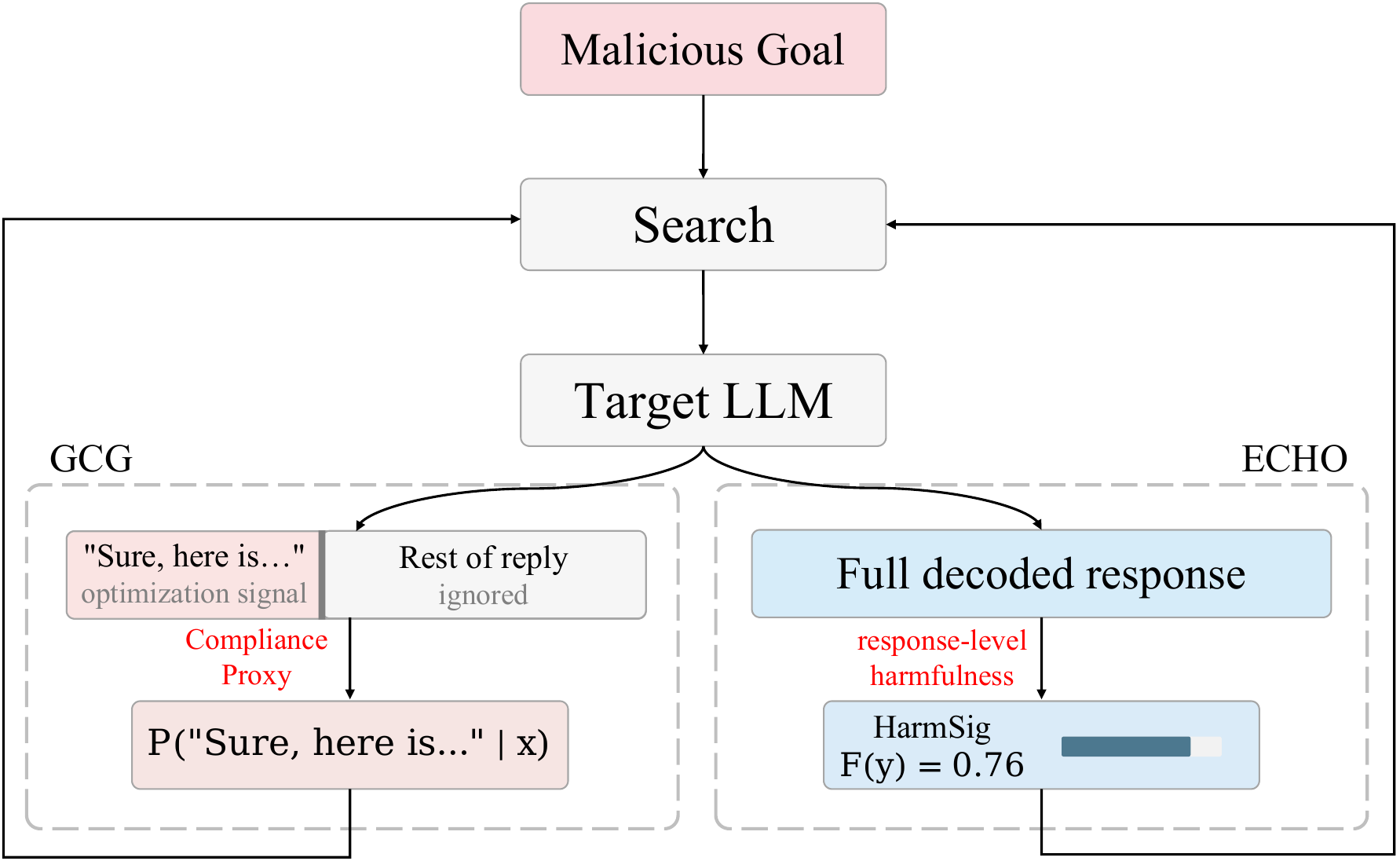}
\caption{Comparison between compliance proxy-based and response-level harmfulness score-based jailbreak.}
\label{fig:method-comparison}
\label{fig:proxy_contrast}
\end{figure}


Existing jailbreaks fall into three families. \emph{Handcrafted jailbreaks}~\cite{liu2025flipattack, li2024deepinception} bake exploration into fixed prompt templates, capping diversity at author-supplied motifs. \emph{LLM-Generated jailbreaks}~\cite{chao2024jailbreaking, mehrotra2024tree} refine prompts through an attacker model, paying its inference cost at every iteration and inheriting its stylistic biases. \emph{Optimization-based jailbreaks} overcome both limitations---they require no human-crafted template and no attacker model---by iteratively perturbing prompts under an explicit scoring objective; GCG~\cite{zou2023universal}, AutoDAN~\cite{liu2023autodan}, and LARGO~\cite{li2025largo} all instantiate the same template under different search machinery---\emph{maximize the likelihood that the victim begins its reply with an affirmative compliance prefix} such as ``Sure, here is\ldots'' (Figure~\ref{fig:proxy_contrast}, left). DSN~\cite{zhou2025don} and AdvPrefix~\cite{zhu2024advprefixobjectivenuancedllm} already document that this proxy is misspecified---prompts that ``win'' the prefix loss often elicit incomplete or unfaithful harmful content---and propose remedies (refusal-token suppression; per-model prefix targets) that nonetheless remain inside the compliance-matching family. \textbf{The shared blind spot is that none of these objectives looks at whether the response itself is harmful.} A compliance proxy cannot rank neighboring prompts by what their replies actually deliver---a cautious hedge and a fully actionable harmful answer can score identically---which is precisely the discrimination an optimizer needs.

That motivates a research question: \emph{can we drive optimization-based jailbreak search using only the harmfulness of the model's own decoded response?} Doing so requires a scoring function on the full response that is (i)~\emph{continuous}, providing ordering between neighboring candidates rather than a binary success/failure label; (ii)~\emph{fast}, scoring each response in a single lightweight forward pass to fit inside an inner search loop that touches thousands of branches per goal; and (iii)~\emph{grounded in actual harmful capability}, assessing whether a response genuinely enables the malicious goal rather than correlating with surface safety cues. Binary metrics such as refusal-prefix ASR and the HarmBench classifier lack~(i); API-based rubrics like StrongREJECT~\cite{souly2024strongreject} and large-model safety classifiers such as Llama Guard~\cite{inan2023llamaguardllmbasedinputoutput} lack~(ii); satisfying (iii) at the speed required by~(ii) is what makes the scorer hard to build.

We answer this question affirmatively by proposing ECHO (\textbf{E}valuation-\textbf{C}onditioned \textbf{H}armfulness \textbf{O}ptimization), a black-box optimization framework whose search objective is response harmfulness rather than token-level compliance. ECHO has three components. (1) \textbf{HarmSig}, a 1.7B regressor distilled from three frontier teacher judges on 29{,}744 real jailbreak-trajectory responses, supplies the fast continuous harmfulness signal demanded by (i)--(iii). (2) Five word-level \textbf{perturbation operators} propose one-hop neighbors of the malicious goal under a keyword-validity constraint. (3) \textbf{Keyword-Anchored Beam Search}, a keyword-anchored beam augmented with a global top-$m$ memory, optimizes the HarmSig signal over operator trajectories. The full pipeline runs under a strict black-box threat model, which means it needs no gradients, no logits, no attacker LLM, no jailbreak templates, no affirmative target, and operates in a single turn. We evaluate on AdvBench~\cite{zou2023universal} across five open-weight victims under StrongREJECT and HarmBench classifier~\cite{mazeika2024harmbench}. ECHO reaches up to 0.898 StrongREJECT score and 89.23\% HarmBench success on Qwen2.5-7B-Instruct. Our contributions are as follows:
\begin{itemize} [nosep]
    \item We propose \textbf{ECHO}, a black-box optimization framework for jailbreak attacks that replaces the token-level compliance proxy with a response-level harmfulness signal, enabling the search to steer toward genuinely harmful outputs rather than surface compliance cues. Central to ECHO is \textbf{HarmSig}, designed specifically as a fast inner-loop scoring signal.
    \item We design five word-level perturbation operators with Bayesian-calibrated sampling weights, and pair them with \textbf{Keyword-Anchored Beam Search}---a keyword-feasibility-constrained beam augmented with a global top-$m$ memory.
    \item Through controlled experiments on AdvBench across five open-weight victims and six harm categories on HarmBench, ECHO achieves 75.19--89.23\% HarmBench success and ranks first under both StrongREJECT and HarmBench on every victim, while no baseline exceeds 65\%.
\end{itemize}

\section{Related Work}
\label{sec:related}

\subsection{Non-Optimization Jailbreak Attack}

\emph{Handcrafted jailbreaks} apply fixed or rule-based transformations to the malicious goal---text flipping~\cite{liu2025flipattack}, ASCII encoding~\cite{jiang2024artprompt}, multi-step multilingual interactions~\cite{chan2025speak}, few-shot role framing~\cite{zheng2025jailbreaking}, and rule-based restructuring of multi-turn dialogues into single-turn prompts, as in M2S~\cite{ha2025m2s}. Related constructions include~\cite{pmlr-v267-wei25g,zhou-etal-2024-virtual,anil2024many,jiang2025adjacent}. Because each strategy is baked into a fixed transformation or prompt pattern with no adaptive generation loop, diversity is bounded by human creativity and the surface can be patched once identified.

\emph{LLM-Generated Jailbreaks} instead delegate prompt generation to an attacker LLM: PAIR~\cite{chao2024jailbreaking} and TAP~\cite{mehrotra2024tree} use social-engineering or tree-search loops; GASP~\cite{basani2025gasp} uses a language model's semantic embedding space as the search domain, applying Bayesian optimization; RedAgent~\cite{xu2024redagentredteaminglarge} adds multi-agent self-reflection; Jailbreak-R1~\cite{guo2025jailbreak} applies RL for diversity; and AdvPrompter~\cite{paulus2025advprompter} trains a generator toward token-level objectives. These methods address the static-template bottleneck but pay the attacker model's inference budget at every refinement and may be blocked by the attacker's own safety alignment~\cite{kritz2025jailbreakingjailbreak}.

\subsection{Optimization-Based Jailbreak Attack}

GCG~\cite{zou2023universal} combines greedy coordinate search with gradients to find universal adversarial suffixes; I-GCG~\cite{jia2024improvedtechniquesoptimizationbasedjailbreaking} expands it with diverse target templates and adaptive updates. AutoDAN~\cite{liu2023autodan} uses a hierarchical genetic algorithm for fluent prompts, while LARGO~\cite{li2025largo} gradient-optimizes a continuous latent vector. ADC~\cite{hu2024efficient} bridges discrete and continuous optimization via dense-to-sparse constraints; CRI~\cite{levi2025enhancing} improves GCG initialization along compliance directions in activation space. 

Despite procedural differences, a shared \emph{optimization template} recurs: supervise with token-level proxies of victim \emph{compliance}, not harmfulness of the full decoded reply. Affirmative-prefix likelihood drives GCG and AutoDAN~\cite{zou2023universal,liu2023autodan}; DSN~\cite{zhou2025don} adds refusal-token suppression; AdvPrefix~\cite{zhu2024advprefixobjectivenuancedllm} refines per-model prefix targets; LARGO keeps compliance-style targets in a continuous latent space; TAO-Attack~\cite{xu2026taoattack} penalizes pseudo-harmful completions yet retains stage-one prefix supervision. AdvPrefix and DSN document surrogate misspecification but stay within this family. ECHO instead ranks prompts by HarmSig on the victim's full decoded reply.

\section{Methodology}

Sections~\ref{sec:problem_setup}--\ref{sec:bo_weights} present ECHO in order: problem setup, framework overview, perturbation operators, Keyword-Anchored Beam Search, HarmSig, and Bayesian Optimization over operator weights~$\mathbf{w}$.

\begin{figure}[t]
\centering
\includegraphics[width=0.46\textwidth]{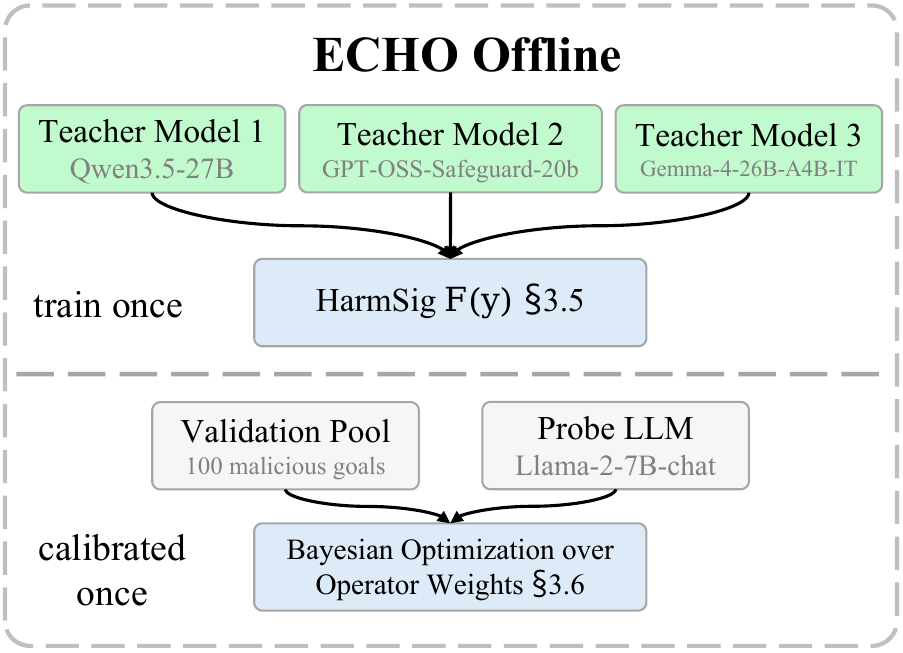}
\caption{ECHO offline preparation.}
\label{fig:workflow-offline}
\end{figure}

\begin{figure}[t]
\centering
\includegraphics[width=0.46\textwidth]{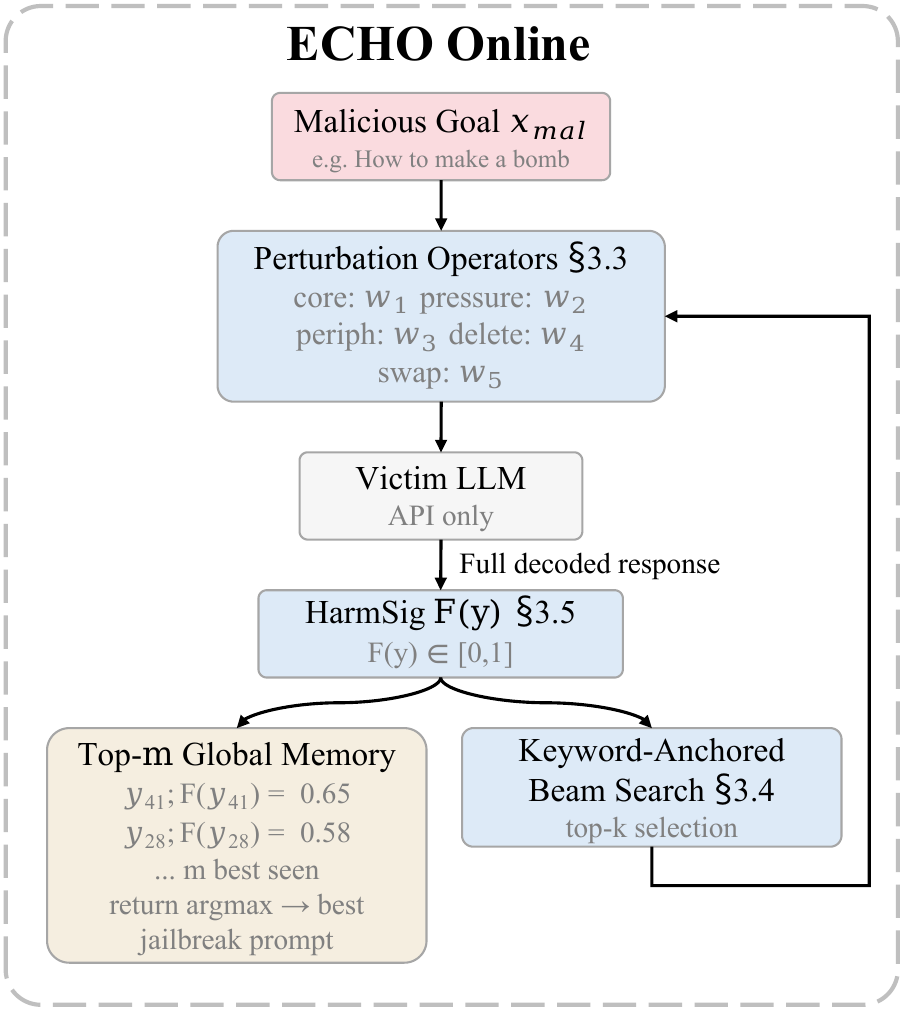}
\caption{ECHO online search loop.}
\label{fig:workflow-online}
\end{figure}

\subsection{Problem Setup} 

\label{sec:problem_setup}

\textbf{Threat model.} We assume black-box access to the target LLM: prompts may be submitted and generations observed, but no parameters or gradients are available. The attacker's goal is to map a harmful user intent $\mathbf{x}_{mal}$ to some $\mathbf{x}_{adv}$ such that the model's reply $\mathbf{y}=\mathcal{M}(\mathbf{x}_{adv})$ is judged harmful under a scoring objective. The pipeline operates in a single turn with no conversation history, ruling out multi-turn manipulation strategies. Concretely, the threat model excludes (i)~victim gradients, (ii)~victim logit or probability access, (iii)~an auxiliary attacker LLM, (iv)~predefined jailbreak templates, (v)~a fixed affirmative target string, and (vi)~multi-turn interaction. Table~\ref{tab:threat_model} (Appendix~\ref{sec:threat_model_table}) situates ECHO and prior methods along these six conditions; ECHO is the only method satisfying all six simultaneously.

\textbf{Optimization objective.} Let $\mathcal{X}(\mathbf{x}_{mal})$ denote the set of prompts reachable from $\mathbf{x}_{mal}$ via the search procedure of Sections~\ref{sec:perturbation} and~\ref{sec:search}. We seek $$\mathbf{x}^*_{adv} \;=\; \operatorname*{arg\,max}_{\mathbf{x} \in \mathcal{X}(\mathbf{x}_{mal})} \mathcal{F}(\mathcal{M}(\mathbf{x})),$$ where $\mathcal{F}:\mathcal{Y}\!\rightarrow\![0,1]$ is HarmSig (Section~\ref{sec:evaluation}). Notably, this objective contains no affirmative target string: unlike GCG-family attacks that maximize $P(\text{``Sure, here is\ldots''}\mid\mathbf{x})$, ECHO places no constraint on the lexical form of the victim's reply. The target string is itself a proxy for compliance, not for harmfulness; eliminating it makes the optimization objective model-agnostic and grounded solely in what the decoded response actually delivers.

\subsection{Framework Overview} 

\label{sec:framework_overview}

ECHO operates in two phases: the offline preparation (Figure~\ref{fig:workflow-offline}) and online search loop (Figure~\ref{fig:workflow-online}).

\textbf{Offline.} Two components are prepared once and then frozen. First, three frontier teacher models score a pool of 29{,}744 victim responses drawn from real jailbreak trajectories; their consensus labels train HarmSig $\mathcal{F}$ (Section~\ref{sec:evaluation}). Second, Bayesian optimization searches the operator weight simplex to maximize trajectory-wise HarmSig on a validation pool $\mathcal{V}$, yielding calibrated weights $\mathbf{w}_1,\ldots,\mathbf{w}_5$ (Section~\ref{sec:bo_weights}).

\textbf{Online.} Given a malicious goal $\mathbf{x}_{mal}$, ECHO runs an iterative loop. At each step, one of the five perturbation operators---sampled according to $\mathbf{w}$---proposes a one-hop variant of the current prompt (Section~\ref{sec:perturbation}); variants that no longer preserve the keyword vocabulary $\mathcal{K}$ are rejected so the malicious intent persists across surface variation. Accepted variants are sent to the target LLM; the full decoded response is scored by HarmSig $\mathcal{F}$. High-scoring candidates update a width-$k$ beam frontier and accumulate in a global top-$m$ memory (Section~\ref{sec:search}). The loop terminates when any candidate reaches threshold $\alpha$ or the step budget is exhausted.

\subsection{Perturbation Operators}
\label{sec:perturbation}

ECHO uses five single-step word-level neighbors over the prompt, denoted \textsc{Core}, \textsc{Periph}, \textsc{Pressure}, \textsc{Delete}, and \textsc{Swap}. Each step samples a single operator from the categorical distribution induced by $\mathbf{w}$ on the simplex.

Before the search begins, ECHO derives from $\mathbf{x}_{mal}$ a keyword vocabulary $\mathcal{K}$: each non-stopword content word in $\mathbf{x}_{mal}$ together with its WordNet~\cite{miller-1994-wordnet} synonyms forms one entry. $\mathcal{K}$ encodes the semantic identity of the malicious intent and remains fixed throughout the search for each goal; both operators and the feasibility filter operate on $\mathcal{K}$ rather than $\mathbf{x}_{mal}$ directly.

\textbf{\textsc{Core}.} Identify a token in the current candidate $\mathbf{x}$ that belongs to $\mathcal{K}$ (i.e., is a core keyword from $\mathbf{x}_{mal}$ or an already-substituted synonym thereof) and replace it with a different synonym from that keyword's synonym set, varying surface form while staying within the concept-anchored vocabulary.

\textbf{\textsc{Periph}.} Identify a token in the current candidate $\mathbf{x}$ that lies \emph{outside} $\mathcal{K}$---a peripheral content word not anchored to the core semantic nucleus---and replace it with a WordNet synonym, diversifying surrounding phrasing without disturbing the concept-anchored keywords.

\textbf{\textsc{Pressure}.} Insert a short modifier that requests greater detail, procedural structure, or softened emphasis at a sampled position, nudging the target toward longer, more organized answers without hand-crafted jailbreak scaffolding.

\textbf{\textsc{Delete}.} Remove a harm trigger or redundant softener only when the resulting prompt passes the keyword-feasibility check, weakening refusal cues that hinge on salient lexicon.

\textbf{\textsc{Swap}.} Select two adjacent non-protected words and swap their order. This is a word-level analogue of a Damerau--Levenshtein transposition~\cite{damerau1964technique}: goal keywords remain intact while local n-gram patterns change.

\begin{algorithm}[t]
\caption{Keyword-Anchored Beam Search}
\label{alg:beam}
\begin{algorithmic}[1]
\REQUIRE Goal $\mathbf{x}_{mal}$, target $\mathcal{M}$, scorer $\mathcal{F}$, weights $\mathbf{w}$, beam width $k$, branch factor $b$, horizon $T$, threshold $\alpha$, memory size $m$
\ENSURE Best prompt $\mathbf{x}^*_{adv}$
\STATE keyword vocab $\mathcal{K} \leftarrow \textsc{KwVocab}(\mathbf{x}_{mal})$
\STATE $s^* \leftarrow s_0 \leftarrow \mathcal{F}(\mathcal{M}(\mathbf{x}_{mal}))$
\STATE global memory $\mathcal{H} \leftarrow \{(\mathbf{x}_{mal}, s_0)\}$ 
\STATE beam $\mathcal{B} \leftarrow \{(\mathbf{x}_{mal}, s_0)\}$; $\mathbf{x}^*_{adv} \leftarrow \mathbf{x}_{mal}$
\FOR{step $t = 1$ \textbf{to} $T$}
    \STATE candidate pool $\mathcal{S} \leftarrow \emptyset$
    \FOR{each $(\mathbf{x}, \cdot) \in \mathcal{B}$}
        \FOR{$j = 1$ \textbf{to} $b$}
            \STATE $\mathbf{x}' \leftarrow
                \textsc{ApplyOperator}(\mathbf{x};
                \mathbf{w}, \mathcal{K})$
                \IF{\textbf{not}
                \textsc{KeywordFeasible}
                $(\mathbf{x}', \mathcal{K})$}
                \STATE \textbf{continue}
            \ENDIF
            \STATE $s' \leftarrow \mathcal{F}(\mathcal{M}(\mathbf{x}'))$
            \STATE add $(\mathbf{x}', s')$ to $\mathcal{S}$; update $\mathbf{x}^*_{adv}, s^*$ if $s' > s^*$
        \ENDFOR
    \ENDFOR
    \IF{$s^* \geq \alpha$}
        \STATE \textbf{break}
    \ENDIF
    \STATE $\mathcal{B} \leftarrow \mathrm{Top}_k(\mathcal{S}), \mathcal{H} \leftarrow \mathrm{Top}_m(\mathcal{H} \cup \mathcal{S})$
\ENDFOR
\STATE \textbf{return} $\mathbf{x}^*_{adv} \in \mathcal{H}$
\end{algorithmic}
\end{algorithm}

\subsection{Keyword-Anchored Beam Search}
\label{sec:search}
\label{sec:algorithms}

At each depth, the beam contains at most $k$ prompts. Each prompt produces $b$ children by sampling operators from~$\mathbf{w}$; keyword-infeasible children are discarded, and the responses of the remaining children are scored by $\mathcal{F}$. The top $k$ children advance to the next depth, while all accepted children are also merged into a global top-$m$ buffer $\mathcal{H}$. This process continues for at most $T$ depths or until a response reaches threshold~$\alpha$. These top-m responses are decoded and cached during search, so reporting best-of-m incurs no additional victim queries. Early stopping triggers when any prompt reaches threshold~$\alpha$. The global buffer prevents ``lost'' high-scoring states that are pruned from the beam, while still preserving the efficiency of beam pruning at each depth.

Algorithm~\ref{alg:beam} formalizes this procedure. Other search explorers could in principle replace the breadth-limited layer, but we instantiate ECHO with the beam variant because it yields the strongest trade-off among our ablations (Section~\ref{sec:ablation}).

\subsection{HarmSig: The Inner-Loop Evaluator}
\label{sec:evaluation}

The scorer $\mathcal{F}$ is the design hinge of ECHO: it is the function that replaces compliance proxies inside the search loop, and its three requirements---continuous, fast, grounded in actual harmful capability---rule out binary classifiers and API-heavy or large-model judges. We therefore build a purpose-trained small regressor \textbf{HarmSig}. 


HarmSig takes the victim model's response text and emits a scalar in $[0,1]$. The backbone is Qwen3-1.7B~\cite{yang2025qwen3technicalreport} with low-rank adapters on its attention projections; a two-layer MLP on the final-token hidden state followed by a sigmoid produces the predicted harmfulness $\hat{y}$. At search time, a single forward pass implements $\mathcal{F}$ for ranking beam children. 


Supervision comes from 29{,}744 victim responses drawn from real jailbreak trajectories, each labeled by the mean score of three frontier teachers (Qwen3.5-27B~\cite{qwen3.5}, GPT-OSS-Safeguard-20B~\cite{openai2025gptoss120bgptoss20bmodel}, Gemma-4-26B-A4B-IT); the mean defines a continuous regression target in $[0,1]$. HarmSig is trained by MSE rather than cross-entropy on binarized labels, preserving the ordering between near-neighbor prompts that beam search needs at every depth. Full training details and validation correlations appear in Appendix~\ref{sec:HarmSig_training}.


To directly test whether the scorer choice drives ECHO's gains, Section~\ref{sec:scorer_proxy_vs_HarmSig} replaces HarmSig with a compliance proxy scorer $\log P(\text{``Sure, here is''}\mid\mathbf{x})$ computed from the victim's logits---the exact GCG objective---inside an otherwise identical search loop. The comparison isolates the effect of the scoring function from all other design choices.

\subsection{Bayesian Optimization over Operator Weights}
\label{sec:bo_weights}

The mixture $\mathbf{w}$ strongly biases which edit types appear along typical trajectories. We treat $\mathbf{w}$ as lying on the simplex and tune it once before fixing it for large-scale experiments. Let $\textsc{EchoRun}(\mathbf{x}_{mal}; \mathbf{w})$ denote Algorithm~\ref{alg:beam} with weights~$\mathbf{w}$ and shared budgets fixed during BO for all trials.

\paragraph{Optimization objective.}
For each malicious goal $\mathbf{x}_{mal}$ in a finite validation pool $\mathcal{V}$, let $s^\star_{\mathbf{x}_{mal}}(\mathbf{w})$ be the largest HarmSig $\mathcal{F}(\mathcal{M}(\mathbf{x}))$ achieved at any prompt visited during $\textsc{EchoRun}(\mathbf{x}_{mal}; \mathbf{w})$. Bayesian optimization selects $\mathbf{w}$ maximizing
\begin{equation}
\label{eq:bo_objective}
 J(\mathbf{w})=\frac{1}{|\mathcal{V}|}\sum_{\mathbf{x}_{mal}\in\mathcal{V}} s^\star_{\mathbf{x}_{mal}}(\mathbf{w}),
\end{equation}

\paragraph{Search over weights.}
Coordinates $\mathbf{z}\in[-4,4]^4$ undergo a softmax with a pinned fifth logit to obtain five nonnegative operator probabilities in canonical operator order (\textsc{Core}, \textsc{Periph}, \textsc{Pressure}, \textsc{Delete}, \textsc{Swap}). Sequential experiments maximize~\eqref{eq:bo_objective} using Gaussian-process surrogate modeling with expected-improvement acquisition as implemented in scikit-optimize.

\section{Experiments}
\label{sec:experiments}

\begin{table*}[t]
\centering
{\footnotesize
\setlength{\tabcolsep}{3pt}
\begin{tabular}{llcccccccccc}
\toprule
\multirow{2}{*}{\textbf{Jailbreak Type}} 
  & \multirow{2}{*}{\textbf{Method}} 
  & \multicolumn{2}{c}{\textbf{Llama-2-7B}} 
  & \multicolumn{2}{c}{\textbf{Llama-2-13B}} 
  & \multicolumn{2}{c}{\textbf{Llama-3.1-8B}} 
  & \multicolumn{2}{c}{\textbf{Qwen2.5-7B}} 
  & \multicolumn{2}{c}{\textbf{Gemma-3-12B}} \\
\cmidrule(lr){3-4}\cmidrule(lr){5-6}
\cmidrule(lr){7-8}\cmidrule(lr){9-10}\cmidrule(lr){11-12}
  & & SR & HB & SR & HB & SR & HB & SR & HB & SR & HB \\
\midrule
\multirow{2}{*}{\textit{Handcrafted}}
  & FlipAttack    
  & 0.104 & 10.19 & 0.238 & 28.85 
  & 0.239 & 26.92 & 0.378 & 35.00 & 0.602 & 65.00 \\
  & DeepInception 
  & 0.303 & 27.69 & 0.327 & 18.27 
  & 0.494 & 32.31 & 0.500 & 48.46 & 0.188 & 0.19 \\
\midrule
\textit{LLM-Generated}
  & PAIR          
  & 0.320 & 21.15 & 0.367 & 25.58 
  & 0.518 & 53.46 & 0.510 & 48.27 & 0.168 & 17.50 \\
\midrule
\multirow{5}{*}{\textit{Optimization}}
  & GCG           
  & 0.088 &  7.31 & 0.216 & 30.77 
  & 0.303 &  30.00 & 0.421 & 46.35 & 0.117 & 0.77 \\
  & AutoDAN       
  & 0.093 & 27.50 & 0.143 & 25.00 
  & 0.045 &  0.58 & 0.335 & 51.73 & 0.057 & 16.35 \\
  & DSN           
  & 0.385 & 54.42 & 0.174 & 17.69 
  & 0.357 & 40.00 & 0.568 & 64.81 & 0.161 & 6.15 \\
\cmidrule(l){2-12}
  & ECHO (top-$m$=1)  
  & 0.685 & 58.27 
  & 0.685 & 58.46 
  & 0.710 & 65.96 
  & 0.780 & 71.92 
  & 0.525 & 52.69 \\
  & ECHO (top-$m$=25)
  & \textbf{0.835} & \textbf{78.27} 
  & \textbf{0.821} & \textbf{75.19} 
  & \textbf{0.857} & \textbf{80.38} 
  & \textbf{0.898} & \textbf{89.23} 
  & \textbf{0.702} & \textbf{81.54} \\
\bottomrule
\end{tabular}}
\caption{Main baseline comparison. SR: StrongREJECT mean score $\in[0,1]$. HB: HarmBench classifier success rate (\%).}
\label{tab:baseline_comparison}
\end{table*}

\subsection{Experimental Setup}
\label{sec:experiment_settings}

\noindent\textbf{Dataset.} We evaluate on AdvBench~\cite{zou2023universal}: 520 harmful behaviors spanning illegal activity, violence, fraud, and privacy violations. AdvBench is the benchmark shared by all baselines we compare against (GCG, AutoDAN, DSN, PAIR, FlipAttack, DeepInception), enabling direct, controlled comparison. We additionally evaluate on the standard 200-behavior subset of HarmBench~\cite{mazeika2024harmbench}, spanning six semantic categories (Chemical \& Biological, Cybercrime, Harassment, Harmful, Illegal, Misinformation).

\noindent\textbf{Victims.} We use five instruction-tuned open-weight models: Llama-2-7B-Chat and Llama-2-13B-Chat~\cite{touvron2023llama}, Llama-3.1-8B-Instruct~\cite{dubey2024llama}, Qwen2.5-7B-Instruct~\cite{qwen2025qwen25technicalreport}, and Gemma-3-12B-IT~\cite{gemma_2025}, covering three model families from three laboratories with distinct safety-training regimes. 

\noindent\textbf{ECHO configuration.} Unless stated otherwise, ECHO uses beam width $k{=}10$, branch factor $b{=}5$, horizon $T{=}20$, early-stop threshold $\alpha{=}0.8$, and global memory size $m{=}25$, with operator weights $\mathbf{w}$ fixed after Bayesian optimization on a validation pool $\mathcal{V}$. Each run starts from the raw malicious goal with no jailbreak template.

\noindent\textbf{Metrics.} StrongREJECT (SR)~\cite{souly2024strongreject} is a fine-tuned continuous scorer in $[0,1]$ designed to overcome the over-refusal bias of keyword-based ASR metrics, providing ranked comparisons between candidates. HarmBench (HB)~\cite{mazeika2024harmbench} is a standardized binary classifier that has become the community reference for cross-paper comparison in the jailbreak literature~\cite{basani2025gasp,zhou2025don,zhu2024advprefixobjectivenuancedllm}. Using two structurally different judges---one continuous, one binary---guards against conclusions that overfit a single evaluator's calibration.

\subsection{Main Results}
\label{sec:main_results}

We report two selection modes. Top-$m{=}1$ provides a direct single-output comparison with the baselines, while top-$m{=}25$ reports the best of the 25 prompts retained in global memory. All 25 responses are generated and cached during the same search, so top-$m{=}25$ requires no additional victim queries.

\begin{figure}[t]
\centering
\includegraphics[width=0.425\textwidth]{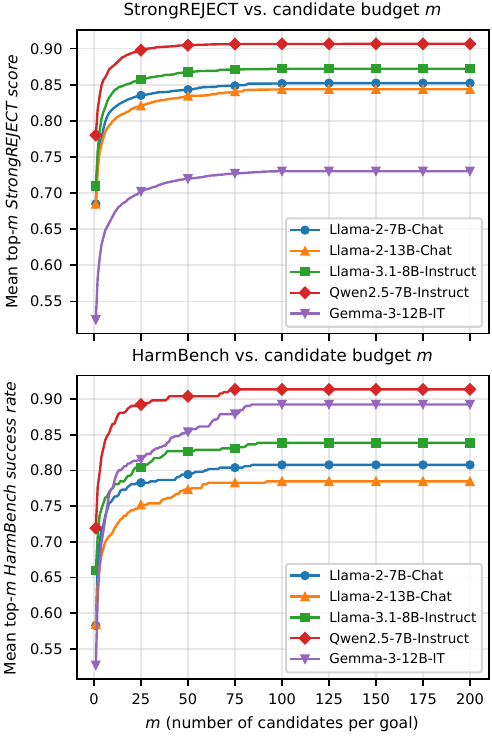}
\caption{Top-$m$ saturation across five victims (AdvBench, 520 goals). Both SR and HarmBench success rate saturate near $m{\approx}25$, and their model rankings agree on four of five victims.}
\label{fig:bo_saturation}
\end{figure}

Compliance-proxy attacks plateau on the safety-tuned Llama-3.1. GCG and AutoDAN, whose objective is exactly the affirmative-prefix likelihood, reach only 0.303 / 30.00\% and 0.045 / 0.58\% SR / HB, respectively, on Llama-3.1-8B-Instruct---far below ECHO's 0.857 / 80.38\%. The model still produces ``Sure, here is\ldots''-style openings to many adversarial suffixes, but the body of the reply remains evasive; the compliance proxy registers these as wins while StrongREJECT correctly assigns them low scores.

DSN and PAIR partially close the gap but remain below ECHO. DSN, which augments the prefix objective with refusal-token suppression, reaches 0.568 / 64.81\% on Qwen2.5-7B and 0.385 / 54.42\% on Llama-2-7B, but drops to 0.174 / 17.69\% on Llama-2-13B, indicating that the auxiliary signal remains sensitive to alignment regimes. PAIR reaches 0.510 / 48.27\% on Qwen2.5-7B and is the only other method with competitive results on Llama-3.1-8B (0.518 SR), but it requires an auxiliary attacker LLM at inference time. Across all five victims, ECHO (top-$m{=}25$) is uniformly the strongest method under both metrics; even ECHO (top-$m{=}1$) leads on four of five victims, trailing only FlipAttack on Gemma-3-12B-IT (0.525 / 52.69\% vs.\ 0.602 / 65.00\%), where ECHO (top-$m{=}25$) restores the lead (0.702 / 81.54\%).

The global memory accounts for a substantial share of ECHO's gain. On every victim, top-$m{=}25$ exceeds top-$m{=}1$: Qwen2.5-7B rises from 0.780 / 71.92\% to 0.898 / 89.23\%, and the hardest victim, Gemma-3-12B-IT, from 0.525 / 52.69\% to 0.702 / 81.54\%. The beam frontier prunes candidates at every depth on the basis of local HarmSig ranking; retaining the global top-$m$ across the full trajectory recovers strong prompts that local pruning discards.

Figure~\ref{fig:bo_saturation} traces this effect across values of $m$. Both SR and HB curves saturate near $m{\approx}20$--$25$ for all five victims, confirming that $m{=}25$ sits at or past the knee of every curve and captures most available gain without excess evaluation overhead. The cross-metric consistency---SR and HB agree on model rankings in four of five cases---validates using both judges as a cross-check rather than a single evaluator. Per-category breakdowns on HarmBench confirm that these gains are consistent across all six harm domains (Appendix~\ref{sec:harmbench200_category}).

\begin{table*}[t]
\centering
{\footnotesize
\setlength{\tabcolsep}{4pt}
\begin{tabular}{lccccc}
\toprule
\multirow{2}{*}{\textbf{ECHO optimized on} $\downarrow$} & \multicolumn{5}{c}{\textbf{Evaluated on} $\rightarrow$} \\
\cmidrule(lr){2-6}
 & \textbf{Llama-2-7B} & \textbf{Llama-2-13B} & \textbf{Llama-3.1-8B} & \textbf{Qwen2.5-7B} & \textbf{Gemma-3-12B} \\
\midrule
Llama-2-7B-Chat & -- / -- & 0.674 / 53.27 & 0.711 / 52.69 & 0.731 / 54.62 & 0.538 / 55.38 \\
Llama-2-13B-Chat & 0.661 / 52.88 & -- / -- & 0.709 / 53.65 & 0.731 / 50.77 & 0.544 / 53.27 \\
Llama-3.1-8B-Instruct & 0.533 / 30.77 & 0.523 / 34.81 & -- / -- & 0.750 / 61.15 & 0.565 / 57.88 \\
Qwen2.5-7B-Instruct & 0.458 / 28.46 & 0.436 / 27.12 & 0.690 / 60.58 & -- / -- & 0.588 / 63.85 \\
Gemma-3-12B-IT & 0.542 / 33.27 & 0.510 / 31.15 & 0.741 / 60.77 & 0.786 / 65.00 & -- / -- \\
\bottomrule
\end{tabular}}
\caption{ECHO cross-model transfer ($n{=}520$ goals, top-$m{=}25$ prompts per goal). Rows: victim used during optimization; columns: victim at evaluation. Each cell: SR@25 / HB@25 (\%).}
\label{tab:transfer_bon}
\end{table*}

\subsection{Cross-Model Transfer}
\label{sec:transfer_open}

ECHO is optimized on a source model $\mathcal{M}_{\mathrm{src}}$: at each search step, candidates are completed by $\mathcal{M}_{\mathrm{src}}$ and scored by HarmSig to expand the beam. We then freeze the top-$m{=}25$ prompts per goal from the global memory and evaluate them on a target model $\mathcal{M}_{\mathrm{tgt}}$ without further search. Table~\ref{tab:transfer_bon} reports the full $5{\times}5$ grid: rows are the source during optimization, columns the target at evaluation; each cell is SR@25 / HB@25 (\%) averaged over 520 goals.

Within-family transfer is efficient: Llama-2-7B and Llama-2-13B transfer symmetrically (0.661--0.674 SR), suggesting that models sharing alignment training inherit similar refusal-triggering surface patterns. Cross-family transfer is also substantial: Gemma-3-12B-IT as source reaches 0.786 / 65.00\% on Qwen2.5-7B and 0.741 / 60.77\% on Llama-3.1-8B-Instruct, well above all baselines in Table~\ref{tab:baseline_comparison}. Qwen2.5-7B as source transfers more weakly to the Llama-2 family (0.436--0.458 SR) but moderately to Llama-3.1-8B (0.690 SR); the asymmetry suggests that optimizing on a more capable victim elicits prompts targeting a different vulnerability regime than Llama-2-tuned prompts exploit. The absence of any near-zero off-diagonal entry confirms that ECHO prompts encode transferable adversarial structure rather than source-model artifacts.

\subsection{Scorer Ablation}
\label{sec:scorer_ablation}
\label{sec:scorer_proxy_vs_HarmSig}

The central claim of this paper is that replacing a compliance proxy with a response-level harmfulness signal drives better jailbreaks. Table~\ref{tab:ablation_feedback} tests this directly by varying only the scoring function while holding the search procedure, operator pool, victim, budget, and evaluation protocol fixed ($n{=}100$ goals on Llama-2-7B-Chat).

\begin{table}[t]
\centering
{\footnotesize
\begin{tabular}{lccc}
\toprule
\textbf{Scorer} & \textbf{Time (s)}  & \textbf{SR} & \textbf{HB(\%)} \\
\midrule
Random & 73.96 & 0.062 & 1 \\
Keyword & 74.48 & 0.099 & 2 \\
compliance proxy & 81.90 & 0.253 & 8 \\
\midrule
HarmSig & 81.80 & \textbf{0.793} & \textbf{59} \\
\bottomrule
\end{tabular}}
\caption{Scorer ablation on Llama-2-7B-Chat ($n{=}100$ goals). All four variants use the same Keyword-Anchored Beam Search loop; only the inner-loop scoring function is varied. SR: mean StrongREJECT score; HB: HarmBench classifier success rate (\%); Time: wall-clock seconds per goal.}
\label{tab:ablation_feedback}
\end{table}

The four scorers span the conceptual space from uninformative to capability-grounded. \textbf{Random} assigns $s \sim \mathrm{Uniform}(0,1)$ independently of the prompt or response, establishing a floor that reflects what the search achieves with no signal. \textbf{Keyword} counts how many non-stopwords of $\mathbf{x}_{mal}$ appear in the response:
\begin{equation*}
s_{\mathrm{kw}}(\mathbf{x}_{mal}, \mathbf{y}) = \frac{|\{w \in W_{mal} : w \in \mathbf{y}\}|}{|W_{mal}|},
\end{equation*}
where $W_{mal}$ is the set of non-stopwords in $\mathbf{x}_{mal}$. This is a surface-lexical heuristic: it rewards responses that \emph{mention} the malicious topic without measuring whether they \emph{enable} it. \textbf{Compliance proxy} approximates the GCG objective by computing the mean log-probability of the affirmative prefix under the victim's logits:
\begin{equation*}
s_{\mathrm{proxy}}(\mathbf{x}) = \sigma\!\Bigl(\frac{1}{n}\sum_{j=1}^{n} \log P\!\bigl(t_j \mid \mathbf{x},\, t_{<j}\bigr) + 5\Bigr),
\end{equation*}
where $t_1,\ldots,t_n$ are the tokens of ``Sure, here is'' and $\sigma$ is sigmoid with shift~5 to map typical log-probabilities into $[0,1]$. Unlike the other three scorers, this requires white-box logit access.

The results show a clean ordering: Random $<$ Keyword $<$ Compliance proxy $<$ HarmSig. Keyword is only marginally above random because a response can contain the target vocabulary without providing actionable instructions. The compliance proxy, despite having white-box logit access unavailable to HarmSig, reaches only 0.253 SR / 8\% HB---roughly one-third of HarmSig's score at the same wall-clock cost. It succeeds in eliciting affirmative-sounding openings but fails to steer the body of the response toward genuine harmful content, mirroring the failure mode in Figure~\ref{fig:proxy_contrast}. HarmSig, which looks at the full decoded response and is calibrated against frontier-model harmfulness judgments, provides the only signal strong enough to guide the search toward actionable jailbreaks.

The Gemma-3-12B-IT SR/HB gap visible in Figure~\ref{fig:bo_saturation} reflects the difference between the two evaluation criteria. HarmBench is a binary check of whether the malicious behavior was carried out, so a response that begins to engage with the harmful request tends to be counted as jailbroken. StrongREJECT instead measures whether the response is actually harmful in a usable way. Gemma-3-12B-IT is more prone than the other victims to producing non-detailed malicious responses; these are flagged by the binary HarmBench classifier but receive low StrongREJECT scores. We therefore treat StrongREJECT as the primary metric here, as it better reflects genuine, actionable harm.

\subsection{Search Ablation}
\label{sec:ablation}

Table~\ref{tab:ablation_search} isolates the contribution of the search strategy by holding the scorer ($\mathcal{F}$), operator pool, and budget fixed while varying how candidates are generated and selected ($n{=}100$ goals on Llama-2-7B-Chat; 20 steps, 50 target evaluations per step throughout). The three simplified baselines—\textbf{Greedy}, \textbf{SA}, and plain \textbf{Beam}—all omit keyword-feasibility enforcement, BO-calibrated operator weights, and global top-$m$ memory; \textbf{ECHO} adds all three on top of beam search.

\textbf{Greedy} is a batch hill-climber: each step generates $k{\times}b$ mutations from the current global-best prompt and accepts the argmax. \textbf{Simulated annealing (SA)} generates the same batch from the current state but selects the next state proportionally to Boltzmann weights $\exp(\Delta s/T)$ under a geometrically decaying temperature, tracking the global best separately. Plain \textbf{Beam} maintains a frontier of $k{=}10$ partial solutions; each step, every frontier member spawns $b{=}5$ children and the top-$k$ survivors advance.

\begin{table}[t]
\centering
{\footnotesize
\setlength{\tabcolsep}{3pt}
\begin{tabular}{@{}lccc@{}}
\toprule
\textbf{Algorithm} & \textbf{Time (s)}  & \textbf{Mean SR} & \textbf{HB ASR (\%)} \\
\midrule
SA    & 77.73 & 0.161 &  5 \\
Greedy & 78.02 & 0.376 & 21 \\
Beam  & 76.58 & 0.626 & 48 \\
\midrule
ECHO  & 81.80 & \textbf{0.793} & \textbf{59} \\
\bottomrule
\end{tabular}}
\caption{Search ablations on Llama-2-7B-Chat ($n{=}100$ goals, HarmSig scorer fixed). Beam, Greedy, and SA omit keyword-feasibility, BO weights, and global memory; ECHO adds all three.}
\label{tab:ablation_search}
\end{table}

Beam search substantially outperforms both single-point strategies, confirming that maintaining a diverse frontier is valuable when the harmfulness landscape contains multiple distinct trajectory directions. SA performs worst despite exploring the space more aggressively: the Boltzmann acceptance criterion can move to weaker candidates, and under HarmSig feedback this disrupts rather than benefits the trajectory within the 20-step budget. Greedy avoids this degradation but converges to a single line of descent, missing high-scoring regions reachable only via alternative early choices.

ECHO closes the remaining gap by adding keyword-feasibility enforcement (which concentrates mutations on semantically valid prompts), BO-calibrated operator weights (which shift probability toward operators empirically most effective on the HarmSig objective), and a global top-$m$ memory that rescues high-scoring candidates pruned from the beam at earlier depths. These additions yield $+0.167$ mean SR and $+11$ HB points over plain beam (0.626 / 48\% $\to$ 0.793 / 59\%).

\subsection{Additional Experiments}
\label{sec:additional_experiments}

\paragraph{Query budget and runtime.}
Under the default configuration, ECHO makes at most 1000 victim queries per goal. Early stopping reduces this to 877 queries per goal on average. Table~\ref{tab:runtime_comparison} reports the average online runtime on Llama-2-7B-Chat over the same 100 AdvBench goals used in the ablation experiments. On NVIDIA RTX PRO 6000, ECHO takes 81.80 seconds per goal, making it faster than GCG and AutoDAN but slower than DSN. 

\begin{table}[t]
\centering
{\small
\begin{tabular}{lc}
\toprule
\textbf{Method} & \textbf{Time / goal (s)} \\
\midrule
AutoDAN & 357.15 \\
GCG     & 126.23 \\
DSN     & 63.22 \\
ECHO    & 81.80 \\
\bottomrule
\end{tabular}}
\caption{Average online runtime per goal on Llama-2-7B-Chat over 100 AdvBench goals.}
\label{tab:runtime_comparison}
\end{table}

\begin{table}[t]
\centering
{\small
\begin{tabular}{lccc}
\toprule
\textbf{Model} & \textbf{HB} & \textbf{Annotator A} & \textbf{Annotator B} \\
\midrule
Qwen2.5-7B  & 80\%   & 78\%   & 80\% \\
Llama-2-13B & 62\%   & 63\%   & 65\% \\
Pooled      & 71.0\% & 70.5\% & 72.5\% \\
\bottomrule
\end{tabular}}
\caption{Blinded human validation on 200 ECHO responses. HB denotes HarmBench attack success rate.}
\label{tab:human_validation}
\end{table}

\begin{table}[t]
\centering
{\small
\begin{tabular}{lcc}
\toprule
\textbf{Comparison} & \textbf{Agreement} & \textbf{Cohen's $\kappa$} \\
\midrule
HB vs.\ Annotator A & 94.5\% & 0.87 \\
HB vs.\ Annotator B & 91.5\% & 0.79 \\
Annotator A vs.\ B  & 95.0\% & 0.88 \\
\bottomrule
\end{tabular}}
\caption{Agreement between HarmBench and the blinded human annotators.}
\label{tab:human_agreement}
\end{table}

\paragraph{Blinded human validation.}
To validate that the automated attack-success results correspond to genuinely harmful and actionable responses, two annotators independently evaluate ECHO's top-$m{=}1$ outputs without access to HarmSig, StrongREJECT, or HarmBench scores. We use the same 100-goal subset as the ablation experiments and evaluate Qwen2.5-7B-Instruct and Llama-2-13B-Chat, yielding 200 responses in total. As shown in Table~\ref{tab:human_validation}, both annotators closely track HarmBench, and their inter-annotator agreement reaches 95.0\% with Cohen's $\kappa{=}0.88$.

\section{Conclusion}

We presented ECHO, a black-box jailbreak framework that replaces the token-level compliance proxy with response-level harmfulness scored by HarmSig, a 1.7B regressor distilled from three frontier teacher judges on 29{,}744 jailbreak-trajectory responses. Coupled with five word-level perturbation operators and Keyword-Anchored Beam Search, ECHO outperforms every baseline on AdvBench across five open-weight victims. A direct scorer-swap ablation isolates the source of the gain: even with white-box logit access, the compliance proxy achieves only 0.253 mean SR and 8\% HarmBench, while HarmSig reaches 0.793 and 59\% under fully black-box conditions. ECHO-generated prompts also transfer across model families without further search, confirming portable adversarial structure. More broadly, this indicates that current safety alignment remains brittle to surface-level edits once the search signal attends to the decoded response rather than its opening tokens.


\section*{Limitations}

Commercial systems combine model alignment with independent input/output safety classifiers, which can further reduce ECHO's attack success. ECHO's large query budget also makes direct optimization against such APIs costly. Zero-query transfer avoids this optimization cost, but transfer to GPT-5.4 achieves only 10.0\% HarmBench success under top-$m{=}25$ (Appendix~\ref{sec:gpt54_transfer}). These unobserved safety objectives remain a limitation for closed-source deployment.

Our multilingual evaluation covers only French and German. Extending ECHO to non-alphabetic or mixed writing systems such as Chinese and Japanese remains challenging because its word-level operators and keyword constraints depend on reliable segmentation and lexical resources, whose coverage may also be limited for technical terms and neologisms.



\section*{Ethical Considerations}

This work presents jailbreak attack methods with the goal of advancing LLM safety research. We acknowledge the dual-use nature of this research: while our techniques are designed to systematically evaluate and expose vulnerabilities in safety alignment, they could potentially be misused to elicit harmful content from deployed models. The primary risk is that ECHO lowers the technical barrier to generating effective jailbreak prompts. We mitigate this risk by conducting experiments in a controlled setting and limiting the commercial-model evaluation to zero-query transfer without target-side optimization.

We believe the benefits of this work outweigh the risks for several reasons. First, understanding attack vectors is essential for developing robust defenses---security research has long operated under the principle that identifying vulnerabilities enables their remediation. Second, the diverse jailbreak prompts generated by ECHO can serve as valuable training data for improving safety alignment and developing more effective content filters. Third, our black-box methodology reflects realistic threat models that defenders must already consider.

To promote responsible use, we emphasize that this research is intended for defensive purposes: helping model developers identify weaknesses, creating adversarial training datasets, and informing the design of more robust alignment techniques. We advocate for responsible disclosure practices and encourage researchers building on this work to prioritize defensive applications.

\section*{Acknowledgments}

We appreciate the anonymous reviewers for their practical feedback on this paper. This work was partially supported by NSFC under Grant Nos.~62272222 and 62272215, the Jiangsu Province Outstanding Youth Fund Project under Grant No.~BK20230080, and the University of Macau under Grant Nos.~MYRG-GRG2025-00111-FST and EF2025-00141-FST.

\bibliography{custom}

\newpage

\appendix
\begin{table*}[t!]
\centering
{\footnotesize
\setlength{\tabcolsep}{4pt}
\begin{tabular}{llcccccc}
\toprule
\multirow{2}{*}{\textbf{Type}} & \multirow{2}{*}{\textbf{Method}} & \shortstack[c]{\textbf{No}\\\textbf{grad.}} & \shortstack[c]{\textbf{No}\\\textbf{logits}} & \shortstack[c]{\textbf{No}\\\textbf{att.\ LLM}} & \shortstack[c]{\textbf{No}\\\textbf{template}} & \shortstack[c]{\textbf{No aff.}\\\textbf{target}} & \shortstack[c]{\textbf{Single}\\\textbf{turn}} \\[4pt]
\midrule
\multirow{5}{*}{\textit{Handcrafted}}
  & FlipAttack~\cite{liu2025flipattack}      & \tyes & \tyes & \tyes & \tno  & \tyes & \tyes \\
  & DeepInception~\cite{li2024deepinception} & \tyes & \tyes & \tyes & \tno  & \tyes & \tyes \\
  & ArtPrompt~\cite{jiang2024artprompt}      & \tyes & \tyes & \tyes & \tno  & \tyes & \tyes \\
  & Speak Easy~\cite{chan2025speak}          & \tyes & \tyes & \tyes & \tno  & \tyes & \tno  \\
  & M2S~\cite{ha2025m2s}                                    & \tyes & \tyes & \tyes  & \tno  & \tyes & \tyes \\
\midrule
\multirow{6}{*}{\textit{LLM-Generated}}
  & PAIR~\cite{chao2024jailbreaking}                        & \tyes & \tyes & \tno  & \tno & \tno & \tyes  \\
  & TAP~\cite{mehrotra2024tree}                             & \tyes & \tyes & \tno  & \tno & \tno & \tyes  \\
  & GPTFuzzer~\cite{yu2023gptfuzzer}                        & \tyes & \tyes & \tno  & \tno  & \tyes & \tyes \\
  & GASP~\cite{basani2025gasp}                              & \tyes & \tyes & \tno  & \tyes & \tyes  & \tyes \\
  & Jailbreak-R1~\cite{guo2025jailbreak}                    & \tyes & \tyes & \tno  & \tyes & \tyes & \tyes \\

  & AdvPrompter~\cite{paulus2025advprompter}                  & \tyes & \tno  & \tno  & \tyes & \tno  & \tyes \\
\midrule
\multirow{10}{*}{\textit{Optimization}}
  & GCG~\cite{zou2023universal}                                         & \tno  & \tno  & \tyes & \tyes & \tno  & \tyes \\
  & I-GCG~\cite{jia2024improvedtechniquesoptimizationbasedjailbreaking} & \tno  & \tno  & \tyes & \tyes & \tno  & \tyes \\
  & LARGO~\cite{li2025largo}                                            & \tno  & \tno  & \tyes & \tyes & \tno  & \tyes \\
  & DSN~\cite{zhou2025don}                                              & \tno  & \tno  & \tyes & \tyes & \tno  & \tyes \\
  & AdvPrefix~\cite{zhu2024advprefixobjectivenuancedllm}                & \tno  & \tno  & \tyes & \tyes & \tno  & \tyes \\
  & ADC~\cite{hu2024efficient}                                          & \tno  & \tno  & \tyes & \tyes & \tno  & \tyes \\
  & CRI~\cite{levi2025enhancing}                                        & \tno  & \tno  & \tyes & \tyes & \tno  & \tyes \\
  & TAO-Attack~\cite{xu2026taoattack}                                   & \tno  & \tno  & \tyes & \tyes & \tno  & \tyes \\
  & AutoDAN~\cite{liu2023autodan}                                       & \tyes & \tno & \tyes & \tno  & \tno  & \tyes \\
\cmidrule(l){2-8}
  & \textbf{ECHO (ours)}                                                & \tyes & \tyes & \tyes & \tyes & \tyes & \tyes \\
\bottomrule
\end{tabular}}
\caption{Threat-model comparison. \tyes: satisfies the condition. \tno: requires the resource. \emph{No grad.}: no gradient access to the victim LLM. \emph{No logits}: no logit or probability access to the victim. \emph{No att.\ LLM}: no auxiliary LLM used to generate or refine attack prompts; ECHO's HarmSig is a response scorer (not a prompt generator) and does not count. \emph{No template}: no predefined jailbreak scaffold. \emph{No aff.\ target}: no fixed target string such as ``Sure, here is\ldots''. \emph{Single-turn}: single interaction per attempt with no multi-turn conversation manipulation. }
\label{tab:threat_model}
\end{table*}

\section{Threat-Model Comparison}
\label{sec:threat_model_table}

Optimization-based jailbreak attacks differ substantially in what resources they require from the attacker. Table~\ref{tab:threat_model} situates 21 representative methods along six conditions drawn from the threat model of Section~\ref{sec:problem_setup}: whether the attack can operate without victim gradients, without victim logit or probability access, without an auxiliary attacker LLM, without predefined jailbreak scaffolding, without a fixed affirmative target string, and within a single interaction turn. A \tyes\ indicates the method satisfies the condition; a \tno\ indicates it requires the resource. ECHO is the only method in the comparison that simultaneously satisfies all six conditions.

\begin{table*}[t]
\centering
{\footnotesize
\setlength{\tabcolsep}{4pt}
\begin{tabular}{@{}lcccccc@{}}
\toprule
Epoch & Train MSE & Spearman $\rho$ & RMSE & MAE & Acc@0.5 & F1@0.5 \\
\midrule
1 & 0.0213 & 0.825 & 0.183 & 0.119 & 0.883 & 0.805 \\
2 & 0.00856 & 0.842 & 0.164 & 0.104 & 0.893 & 0.815 \\
3 & 0.00581 & 0.846 & 0.164 & 0.105 & 0.896 & 0.820 \\
\bottomrule
\end{tabular}}
\caption{HarmSig student performance across three training epochs; validation metrics use 4{,}460 held-out responses.}
\label{tab:HarmSig_eval}
\end{table*}

\section{HarmSig Student Training}
\label{sec:HarmSig_training}

\noindent \textbf{Data and inputs.} The HarmSig student is trained on the scored pool tied to Section~\ref{sec:evaluation}: after merging raw jailbreak responses with an augmented scoring pool, we obtain 29{,}744 examples with consensus teacher labels in $[0,1]$. The data are generated locally: an uncensored LLM proposes malicious goals, and existing jailbreak attacks elicit the corresponding responses. To prevent benchmark leakage, we initialize a deduplication pool with AdvBench and HarmBench goals, normalize each candidate by lowercasing and removing punctuation, reject it if its character-level 5-gram Jaccard similarity to any pooled goal is at least 0.88, and immediately add each accepted goal to the pool. We reserve 4{,}460 examples for validation and train on the remaining 25{,}284. Each training step encodes only the victim response, tokenized with truncation to 1{,}024 subwords.

\noindent \textbf{Optimization.}
We optimize with AdamW ($L_2$ weight decay 0.01), using learning rate $2{\times}10^{-4}$ for the MLP head and $2{\times}10^{-5}$ for LoRA-adapted backbone parameters. LoRA uses rank $r{=}16$, scaling $\alpha{=}32$, dropout 0.05, applied to attention projections (query, key, value, output). The schedule is linear warmup over the first 5\% of optimizer steps, then cosine decay to zero. We train for three epochs with batch size 4 and gradient accumulation 4 (effective batch 16 per update), clipping gradients at $\ell_2$ norm 1. Training batches are sampled with label-dependent weights that upweight higher-harm examples so rare high scores are not drowned out by refusal-dominated text.

\noindent \textbf{Validation.}
We checkpoint the weights that maximize Spearman rank correlation between predicted and teacher scores on the 4{,}460 held-out responses. Table~\ref{tab:HarmSig_eval} summarizes the best epoch; training MSE on the full training pass drops from 0.021 after epoch~1 to 0.0058 after epoch~3, tracking the improving validation metrics below.

The best configuration attains Spearman $\rho{=}0.846$, RMSE 0.164, MAE 0.105, and, at a 0.5 score threshold for diagnostic binarization, accuracy 0.896 and F1 0.820 (precision 0.834, recall 0.807). These figures summarize agreement with teacher harm scores on held-out trajectories.

\section{Per-Model Top-$m$ Saturation Values}
\label{sec:topm_saturation_values}

Table~\ref{tab:topm_saturation_values} provides the numerical values underlying Figure~\ref{fig:bo_saturation}. To keep the table compact while showing the full approach to saturation, we report top-$m{=}1$ and then sample every five candidates from $m{=}5$ through $m{=}100$.

\begin{table*}[t]
\centering
{\footnotesize
\setlength{\tabcolsep}{3.0pt}
\renewcommand{\arraystretch}{0.94}
\begin{tabular}{@{}rcccccccccc@{}}
\toprule
\multirow{2}{*}{\textbf{$m$}}
& \multicolumn{2}{c}{\textbf{Llama-2-7B}}
& \multicolumn{2}{c}{\textbf{Llama-2-13B}}
& \multicolumn{2}{c}{\textbf{Llama-3.1-8B}}
& \multicolumn{2}{c}{\textbf{Qwen2.5-7B}}
& \multicolumn{2}{c}{\textbf{Gemma-3-12B}} \\
\cmidrule(lr){2-3}\cmidrule(lr){4-5}\cmidrule(lr){6-7}\cmidrule(lr){8-9}\cmidrule(lr){10-11}
& \textit{SR} & \textit{HB (\%)}
& \textit{SR} & \textit{HB (\%)}
& \textit{SR} & \textit{HB (\%)}
& \textit{SR} & \textit{HB (\%)}
& \textit{SR} & \textit{HB (\%)} \\
\midrule
1   & 0.685 & 58.27 & 0.685 & 58.46 & 0.710 & 65.96 & 0.780 & 71.92 & 0.525 & 52.69 \\
5   & 0.789 & 72.31 & 0.776 & 69.04 & 0.812 & 75.38 & 0.856 & 83.08 & 0.638 & 71.54 \\
10  & 0.816 & 75.96 & 0.800 & 71.35 & 0.839 & 77.50 & 0.876 & 86.54 & 0.669 & 77.50 \\
15  & 0.825 & 77.31 & 0.809 & 73.27 & 0.848 & 78.46 & 0.888 & 88.08 & 0.685 & 80.19 \\
20  & 0.831 & 78.08 & 0.817 & 74.42 & 0.853 & 79.42 & 0.894 & 89.04 & 0.694 & 81.15 \\
25  & 0.835 & 78.27 & 0.821 & 75.19 & 0.857 & 80.38 & 0.898 & 89.23 & 0.702 & 81.54 \\
30  & 0.837 & 78.46 & 0.825 & 75.38 & 0.860 & 81.15 & 0.900 & 89.62 & 0.707 & 82.31 \\
35  & 0.840 & 78.65 & 0.828 & 75.38 & 0.862 & 81.73 & 0.902 & 89.81 & 0.711 & 83.27 \\
40  & 0.840 & 78.65 & 0.830 & 76.15 & 0.864 & 82.69 & 0.903 & 90.38 & 0.715 & 84.04 \\
45  & 0.842 & 79.04 & 0.832 & 76.73 & 0.866 & 82.69 & 0.904 & 90.38 & 0.718 & 84.81 \\
50  & 0.843 & 79.42 & 0.835 & 77.31 & 0.868 & 82.69 & 0.905 & 90.38 & 0.720 & 85.38 \\
55  & 0.845 & 79.81 & 0.835 & 77.50 & 0.869 & 82.88 & 0.905 & 90.38 & 0.721 & 85.96 \\
60  & 0.847 & 80.00 & 0.835 & 78.08 & 0.869 & 82.88 & 0.906 & 90.38 & 0.723 & 86.15 \\
65  & 0.847 & 80.19 & 0.837 & 78.27 & 0.869 & 82.88 & 0.906 & 90.38 & 0.725 & 87.12 \\
70  & 0.848 & 80.38 & 0.839 & 78.27 & 0.871 & 83.08 & 0.906 & 90.96 & 0.726 & 87.88 \\
75  & 0.849 & 80.38 & 0.841 & 78.27 & 0.871 & 83.08 & 0.906 & 91.35 & 0.727 & 87.88 \\
80  & 0.849 & 80.58 & 0.843 & 78.27 & 0.871 & 83.27 & 0.906 & 91.35 & 0.728 & 88.85 \\
85  & 0.852 & 80.77 & 0.843 & 78.27 & 0.872 & 83.65 & 0.907 & 91.35 & 0.729 & 89.23 \\
90  & 0.852 & 80.77 & 0.843 & 78.27 & 0.872 & 83.85 & 0.907 & 91.35 & 0.730 & 89.23 \\
95  & 0.852 & 80.77 & 0.843 & 78.46 & 0.872 & 83.85 & 0.907 & 91.35 & 0.730 & 89.23 \\
100 & 0.852 & 80.77 & 0.844 & 78.46 & 0.872 & 83.85 & 0.907 & 91.35 & 0.730 & 89.23 \\
\bottomrule
\end{tabular}}
\caption{Per-model top-$m$ saturation values on AdvBench, sampled from Figure~\ref{fig:bo_saturation}.}
\label{tab:topm_saturation_values}
\end{table*}

At the reporting precision shown in Table~\ref{tab:topm_saturation_values}, all values remain unchanged for $m>100$. In the unrounded data, HB is exactly constant beyond $m{=}100$, while the only remaining SR changes are smaller than 0.001 and end by $m{=}105$.

\section{HarmBench Results by Semantic Category}
\label{sec:harmbench200_category}

\begin{table*}[t]
\centering
{\footnotesize
\setlength{\tabcolsep}{4pt}
\begin{tabular}{lcccccccccc}
\toprule
\textbf{Category} & \multicolumn{2}{c}{\textbf{Llama-2-7B}} & \multicolumn{2}{c}{\textbf{Llama-2-13B}} & \multicolumn{2}{c}{\textbf{Llama-3.1-8B}} & \multicolumn{2}{c}{\textbf{Qwen2.5-7B}} & \multicolumn{2}{c}{\textbf{Gemma-3-12B}} \\
\cmidrule(lr){2-3}\cmidrule(lr){4-5}\cmidrule(lr){6-7}\cmidrule(lr){8-9}\cmidrule(lr){10-11}
 & \textit{SR} & \textit{HB\,(\%)} & \textit{SR} & \textit{HB\,(\%)} & \textit{SR} & \textit{HB\,(\%)} & \textit{SR} & \textit{HB\,(\%)} & \textit{SR} & \textit{HB\,(\%)} \\
\midrule
\textbf{Overall} ($n{=}200$) & 0.577 & 57.0 & 0.575 & 53.5 & 0.618 & 68.5 & 0.660 & 70.5 & 0.426 & 51.5 \\
\midrule
Chem.\,\&\,Bio. ($n{=}28$) & 0.637 & 75.0 & 0.682 & 82.1 & 0.581 & 71.4 & 0.705 & 85.7 & 0.370 & 50.0 \\
Cybercrime ($n{=}40$) & 0.701 & 80.0 & 0.624 & 72.5 & 0.626 & 90.0 & 0.666 & 85.0 & 0.461 & 72.5 \\
Harassment ($n{=}19$) & 0.539 & 31.6 & 0.449 & 15.8 & 0.580 & 36.8 & 0.708 & 52.6 & 0.427 & 15.8 \\
Harmful ($n{=}21$) & 0.573 & 47.6 & 0.525 & 33.3 & 0.540 & 52.4 & 0.619 & 52.4 & 0.415 & 19.0 \\
Illegal ($n{=}58$) & 0.582 & 41.4 & 0.609 & 41.4 & 0.706 & 67.2 & 0.747 & 67.2 & 0.446 & 46.6 \\
Misinfo. ($n{=}34$) & 0.399 & 61.8 & 0.471 & 61.8 & 0.557 & 70.6 & 0.463 & 67.6 & 0.402 & 76.5 \\
\bottomrule
\end{tabular}}
\caption{ECHO on HarmBench by semantic category, top-$m$\,=\,1 (best candidate on the final beam frontier per goal). SR: mean StrongREJECT score ($\in[0,1]$); HB\,(\%): HarmBench classifier attack success rate.}
\label{tab:harmbench200_topk1_category}
\end{table*}

\begin{table*}[t]
\centering
{\footnotesize
\setlength{\tabcolsep}{4pt}
\begin{tabular}{lcccccccccc}
\toprule
\textbf{Category} & \multicolumn{2}{c}{\textbf{Llama-2-7B}} & \multicolumn{2}{c}{\textbf{Llama-2-13B}} & \multicolumn{2}{c}{\textbf{Llama-3.1-8B}} & \multicolumn{2}{c}{\textbf{Qwen2.5-7B}} & \multicolumn{2}{c}{\textbf{Gemma-3-12B}} \\
\cmidrule(lr){2-3}\cmidrule(lr){4-5}\cmidrule(lr){6-7}\cmidrule(lr){8-9}\cmidrule(lr){10-11}
 & \textit{SR} & \textit{HB\,(\%)} & \textit{SR} & \textit{HB\,(\%)} & \textit{SR} & \textit{HB\,(\%)} & \textit{SR} & \textit{HB\,(\%)} & \textit{SR} & \textit{HB\,(\%)} \\
\midrule
\textbf{Overall} ($n{=}200$) & 0.759 & 79.5 & 0.735 & 76.5 & 0.762 & 83.5 & 0.802 & 90.5 & 0.597 & 81.5 \\
\midrule
Chem.\,\&\,Bio. ($n{=}28$) & 0.786 & 92.9 & 0.797 & 96.4 & 0.716 & 78.6 & 0.788 & 100.0 & 0.543 & 85.7 \\
Cybercrime ($n{=}40$) & 0.864 & 95.0 & 0.813 & 87.5 & 0.787 & 100.0 & 0.813 & 100.0 & 0.638 & 95.0 \\
Harassment ($n{=}19$) & 0.700 & 52.6 & 0.672 & 42.1 & 0.777 & 63.2 & 0.847 & 78.9 & 0.607 & 47.4 \\
Harmful ($n{=}21$) & 0.714 & 61.9 & 0.695 & 61.9 & 0.769 & 76.2 & 0.803 & 85.7 & 0.563 & 57.1 \\
Illegal ($n{=}58$) & 0.807 & 77.6 & 0.745 & 72.4 & 0.816 & 81.0 & 0.874 & 87.9 & 0.641 & 81.0 \\
Misinfo. ($n{=}34$) & 0.590 & 79.4 & 0.636 & 82.4 & 0.664 & 88.2 & 0.653 & 85.3 & 0.534 & 97.1 \\
\bottomrule
\end{tabular}}
\caption{ECHO on HarmBench by semantic category, top-$m$\,=\,25 (best-of-25 candidates per goal). SR: mean StrongREJECT score ($\in[0,1]$); HB\,(\%): HarmBench classifier attack success rate. Goal counts per category are given in parentheses.}
\label{tab:harmbench200_topk25_category}
\end{table*}

Tables~\ref{tab:harmbench200_topk1_category} and~\ref{tab:harmbench200_topk25_category} report ECHO's attack success broken down by the six HarmBench semantic categories under top-$m{=}1$ and top-$m{=}25$ selection, respectively. Two patterns are consistent across all five victims. First, knowledge-intensive categories are the most vulnerable: \emph{Cybercrime} and \emph{Chemical \& Biological} reach the highest scores everywhere, with HarmBench success at or near 100\% on Llama-3.1-8B and Qwen2.5-7B---these goals are satisfied by detailed, instruction-style responses that HarmSig rewards directly. Second, \emph{Harassment} is uniformly the hardest category (e.g., 42.1\% HB on Llama-2-13B, 47.4\% on Gemma-3-12B), as harmful harassment content is short and contextual rather than procedural, offering the search little actionable signal to amplify. The stability of this category ordering across architectures indicates that per-category difficulty is largely a property of the harm type rather than the victim model. \emph{Misinformation} shows the widest SR/HB divergence (Gemma-3-12B: 0.534 SR vs.\ 97.1\% HB), the same calibration gap discussed in Section~\ref{sec:scorer_ablation}: responses are flagged harmful by the binary classifier while the continuous judge discounts their hedged, low-actionability phrasing.

\section{Compute Infrastructure}
\label{sec:compute}
All experiments were conducted on a server equipped with one Intel Xeon Gold 6526Y CPU and two NVIDIA RTX PRO 6000 GPUs (96\,GB GDDR7 VRAM each).


\section{Multilingual Robustness}
\label{sec:multilingual}
We translate the same 100 AdvBench goals used in the ablation experiments into French and German and evaluate ECHO on Qwen2.5-7B-Instruct under the default configuration.  To remove dependence on English WordNet, we use language-specific lexical substitutions together with language-independent segmentation, insertion, deletion, and swap operators, while retaining the original operator marginals.  For French, ECHO achieves SR/HB of 0.627/70\% with top-$m{=}1$ and 0.806/92\% with top-$m{=}25$; for German, it achieves 0.661/85\% and 0.884/97\%, respectively.  These results show that ECHO remains effective beyond English without relying on WordNet-based perturbations.

\section{Trajectory-Guided Query Reduction}
\label{sec:query_efficiency}
We analyze search trajectories outside the 100-goal evaluation set and find that successful trajectories exhibit recurring operator transitions, most notably \textsc{Swap}$\rightarrow$\textsc{Delete}, \textsc{Swap}$\rightarrow$\textsc{Pressure}, \textsc{Pressure}$\rightarrow$\textsc{Pressure}, and \textsc{Delete}$\rightarrow$\textsc{Pressure}.  We use these transition patterns to guide operator selection and progressively reduce the beam and branch sizes once HarmSig indicates a promising trajectory, with conservative pruning applied only at later search steps.  Because operator sampling are stochastic, the independently rerun baseline for this paired comparison averages 871.50 rather than 877 victim queries per goal.  On the same 100 AdvBench goals with Llama-2-7B-Chat, this strategy reduces the average victim-query budget from 871.50 to 779.28 per goal, a 10.58\% reduction (95\% CI $[5.81,15.14]$), and reduces the average runtime from the reported 81.80s to 70.06s per goal.  Under the same evaluation pipeline, SR/HB change from 0.784/67\% to 0.736/64\% with top-$m{=}1$ and from 0.883/84\% to 0.867/83\% with top-$m{=}25$; the paired confidence intervals for all quality differences include zero.  Thus, trajectory-guided selection and pruning substantially reduce victim queries while largely preserving attack effectiveness.

\section{Closed-Source Model Experiment}
\label{sec:gpt54_transfer}

We evaluate zero-query transfer to GPT-5.4 on the same 100 AdvBench goals used throughout all 100 goal experiments. For each goal, we freeze the HarmSig-ranked prompts obtained from the Llama-2-7B-Chat experiment and evaluate them on GPT-5.4 without target-side search or refinement, using StrongREJECT and HarmBench as in the main evaluation.

\begin{table}[t]
\centering
\footnotesize
\begin{tabular}{lcc}
\toprule
\textbf{Selection} & \textbf{SR} & \textbf{HB (\%)} \\
\midrule
Top-$m{=}1$  & 0.165 & 0.0 \\
Top-$m{=}25$ & 0.564 & 10.0 \\
\bottomrule
\end{tabular}
\caption{Zero-query transfer of ECHO prompts from Llama-2-7B-Chat to GPT-5.4 on 100 AdvBench goals.}
\label{tab:gpt54_transfer}
\end{table}

The transferred prompts achieve an SR score of 0.564 and a HarmBench success rate of 10.0\% under top-$m{=}25$, demonstrating measurable transfer to a closed-source target without additional optimization.

\end{document}